\documentclass[numberedappendix]{emulateapj}
\slugcomment{Accepted by The Astrophysical Journal}
\shorttitle{Stellar age versus mass of early types}
\shortauthors{Lisker \& Han}
\begin{document}
 
\title{Stellar age versus mass of early-type galaxies in the Virgo Cluster}

\author{Thorsten Lisker}
\affil{Astronomisches Rechen-Institut, Zentrum f\"ur Astronomie der
  Universit\"at Heidelberg (ZAH), M\"onchhofstra\ss e 12-14, D-69120
  Heidelberg, Germany
}
\email{TL@x-astro.net}

\and

\author{Zhanwen Han}
\affil{National Astronomical Observatories\,/\,Yunnan Observatory, The
  Chinese Academy of Sciences, Kunming 650011, China
}
\email{zhanwenhan@hotmail.com}

\begin{abstract}
The flux excess of elliptical galaxies in the far-ultraviolet 
can be reproduced by population synthesis models when accounting for the
population of old hot helium-burning subdwarf stars. This has
been achieved by Han and coworkers through a quantitative model of binary
stellar evolution. Here, we 
compare the resulting evolutionary population synthesis model to the
{\it GALEX} far$-$near ultraviolet colors (FUV$-$NUV) of Virgo cluster early-type
galaxies that were published by Boselli and coworkers.
FUV$-$NUV is reddest at about
the dividing luminosity of dwarf and giant galaxies, and becomes increasingly
blue for both brighter and fainter luminosities. This behavior can be easily
explained by the binary model with a continuous sequence
 of longer duration and later truncation of star formation at lower galaxy
masses. Thus, in contrast to previous conclusions, the {\it GALEX} data do
not require a dichotomy between the stellar population properties of dwarfs
and giants. Their apparently opposite behavior in FUV$-$NUV occurs naturally
when the formation of hot subdwarfs through binary evolution is taken into
account.
\end{abstract}
 


\section{Introduction and Motivation}
 \label{sec:intro}

What can we learn from the ultraviolet light of early-type galaxies that
longer wavelengths cannot tell us?
Mainly from spectrophotometric analyses in the optical and near-infrared,
a picture emerged in which the stars of massive early-type galaxies were
formed during a short period of intense star formation at early epochs
\citep{bow92,ell97,tho99}. On the other hand, 
studies of their stellar mass or number density evolution with redshift imply that a
substantial fraction of them must have formed within the last gigayears
\citep{bell04,ferlis05}.
Both above statements were found to depend on galaxy mass: at lower
masses, star formation lasted longer \citep{nel05,tho05} and the number density shows a
stronger evolution with redshift \citep{cim06}.

While optical spectral line indices are a commonly used tool to derive
estimates for the (mean) ages and metallicities of early-type galaxies,
they still leave room for discussion. Larger values of the $H_\beta$ index
were interpreted by \citet{dej97} with lower mean ages,
caused by the presence of a relatively young stellar population on top of
an old one that dominates the mass. However, \citet{mar00} found that a
large value of $H_\beta$ can alternatively be explained by the contribution
of old metal-poor stars. Nevertheless, \citet{cal03} derived substantially
lower mean ages, as well as a larger age spread, for early-type galaxies
with lower velocity dispersions (i.e., lower masses), even when taking
into account the approach of \citet{mar00}.
 
To what extent recent or residual star formation does play a role in
shaping present-day early-type galaxies can be explored with the 
ultraviolet (UV) data of the {\it Galaxy Evolution Explorer} \citep[{\it
    GALEX}; ][]{galex}, since the contribution of young stars is much
stronger in the UV than at optical wavelengths. From these data,
\citet{kav07} and \citet{sch07} concluded that about 30\% of
massive early types at low redshift have experienced some star
formation less than 1\,Gyr ago.
Similar results were obtained at higher redshifts, for which the
rest-frame UV can be probed with optical surveys: \citet{kav07b} found
``compelling evidence that early-types of all luminosities form stars over
the lifetime of the Universe''. Again, less  massive objects were found to
have formed most of their stellar mass later than more
massive galaxies \citep[also see][]{fersilk00}.

The above mentioned studies, while covering each a range of masses and
luminosities of early-type galaxies, did not reach into the dwarf regime.
Based on a study of {\it GALEX} far$-$near UV (FUV$-$NUV) color of early-type
galaxies in the Virgo cluster, \citet{bos05} reported a pronounced
dichotomy between dwarf and giant early types: while giants become bluer
with \emph{increasing} near-infrared $H$ luminosity, dwarfs do so with
\emph{decreasing} luminosity. The former can be understood with 
a stronger presence of the well-known FUV flux excess (``UV-upturn'')
at higher masses \citep{gil07}, while the latter was interpreted by
\citet{bos05} with residual star formation in early-type dwarfs.
These observations were of particular importance, for they appeared to
settle an issue that has never been conclusively resolved in the optical
regime: the question of whether or not dwarf and giant early-type galaxies
follow the same color-magnitude relation \citep{deV61,sec97,conIII}.

Interpretations of integrated galaxy light hinge critically on
the availability of appropriate stellar population synthesis models
that can be compared to the data. A crucial issue is the availability of a
quantitative prescription for the cause of the FUV excess. 
Commonly used models, which are based on single stellar evolution, 
 do describe a rise in the FUV at old stellar population ages
 \citep[e.g.][]{mar00,bc03}.
However, the approach of \citet{kav07}, who preferred to use the
spectral energy distribution of a single galaxy as empirical template for
the FUV excess, demonstrates the need for further improvement in modelling its
cause.

While it is believed that hot
helium-burning subdwarf B (sdB) stars on the Extreme Horizontal 
Branch \citep[EHB,][]{heb86} and their progeny \citep{str07} are
the main cause of the FUV excess \citep{bro97,bro00},
until recently the formation of these stars was only poorly
understood \citep{heb02,ede03,lis05a}. Since close binary evolution was
known to play a major role in it
\citep{max01}, a binary stellar evolution model for the formation of hot
subdwarfs was developed by \citet{han02,han03}, which is able to account
for the observed range of stellar parameters \citep{lis05a}.
This model was integrated into an evolutionary population synthesis model
by \citet{han07}, which, as we shall demonstrate below, is not
only able to explain the observed UV colors of Virgo cluster early-type
galaxies, but also provides a simple explanation for the
apparent dichotomy between dwarf and giant early types.


\section{The binary population synthesis model}
 \label{sec:model}

\begin{figure}
  \epsscale{1.0}
  \plotone{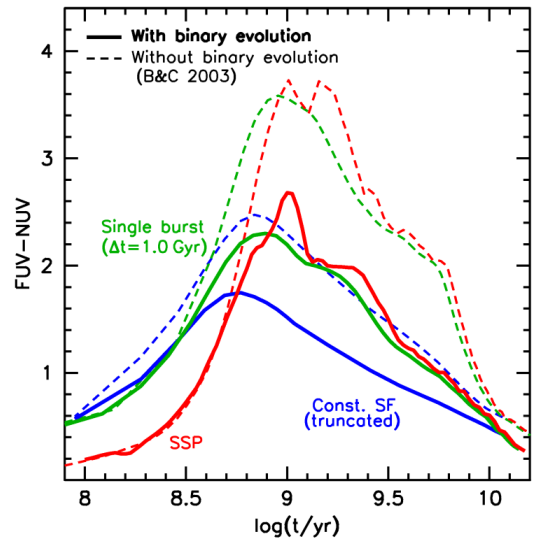}
  \caption{{\bf Binary vs.\ single stellar evolution.}
    Time evolution of FUV$-$NUV color of various synthetic stellar populations,
    computed with the binary model of \citet[solid lines]{han07} and the
    single-star model of \citet[dashed lines]{bc03}. For each of the two
    models, we show the color evolution of  a simple stellar population (SSP,
    red or black), a stellar population formed in a single 1-Gyr period
    (``Single burst'', green or light grey), and a population formed through
    continuous star formation that is truncated at a certain epoch (``Const.\
    SF'', blue or dark grey; see text for details). The ``age'' 
    $t$ is defined for all cases as the time since the end of star formation.
  }
  \label{fig:noobs}
\end{figure}

By incorporating the binary model of \citet{han02,han03} 
for the formation of hot subdwarfs into evolutionary
population synthesis modelling,
Han, Podsiadlowski \& Lynas-Gray (2007, hereafter HPL07) obtained an ``a
priori'' model for the UV-upturn of elliptical galaxies. Their study
indicated that the UV-upturn is most likely the result of binary interactions.
In the model, HPL07 used their binary population synthesis code to 
evolve millions of stars (including binaries) from the zero-age main-sequence
to the white dwarf stage or a supernova explosion. 
The spectra of hot subdwarfs were calculated with the {\scriptsize ATLAS9}
stellar atmosphere code \citep{kur92}, while the spectra of other stars were
taken from the latest version of the comprehensive BaSeL spectral library
\citep{lej97,lej98}. The model simulates the evolution of the colors and
the spectral energy distribution of a simple stellar population (SSP).

The rest-frame FUV$-$NUV color evolution of the SSP
is shown in Fig.~\ref{fig:noobs} (solid red or black
line). Its main characteristic is the rather sharp turnaround at an age of
$t\approx 1$\,Gyr, at which the hot subdwarf stars start to
contribute to the FUV flux and thus cause a subsequent bluening of
FUV$-$NUV. We compare the binary model to the single stellar evolution model
of \citet[hereafter BC03; dashed red or black line]{bc03}, since it is
one of the most commonly used models in stellar population studies of
galaxies.
The BC03 model reaches to much redder FUV$-$NUV colors than does
the HPL07 model, but it eventually also becomes bluer due to 
post-AGB stars \citep{bc03}.
See Sect.~\ref{sec:sub_othermodels} for a comparison to other
models.

From each SSP we computed the evolution of a stellar population
with a star formation history that, compared to the SSP's single
instantaneous burst, is somewhat more realistic: a single star formation period of
finite length, and continuous star formation that is truncated at a certain
epoch. For the first case, we adopt a  
1-Gyr period of constant star formation rate (``Single burst'', green or
light grey lines in 
Fig.~\ref{fig:noobs}).
For the second case, we adopt a 
constant star formation rate since 12.6\,Gyr\footnote{We actually use
  $10^{10.1}\,{\rm yr}=12.59$\,Gyr}
until a point of time when star formation is being truncated (``Const.\
SF'', blue or dark grey lines in Fig.~\ref{fig:noobs}).
In order to have a uniform definition of ``age'' for these different model
populations, we adopt the time $t$ that has passed since the \emph{end} of
star formation as a proxy for age. Actually, $t$ directly indicates the
age of the youngest stellar component, which will be helpful for the
discussion in Sect.~\ref{sec:discuss}.

The above computation was done with the {\it csp\_galaxev} tool of \citeauthor{bc03},
which uses the method of isochrone synthesis \citep[see][]{bc03}. Since the
HPL07 binary model is based on a chemical composition of $X=0.78, Y=0.20, Z=0.02$,
we use the BC03 model with the same composition, based on 'Padova 1994'
isochrones \citep{pad94_1,pad94_2,pad94_3,pad94_4,pad94_5} and adopting a
\citet{cha03} IMF.


\section{Comparison of models and observations}
\label{sec:compare}

\subsection{Binary versus single-star model}
\label{sec:sub_compare}

\begin{figure}
  \epsscale{1.0}
  \plotone{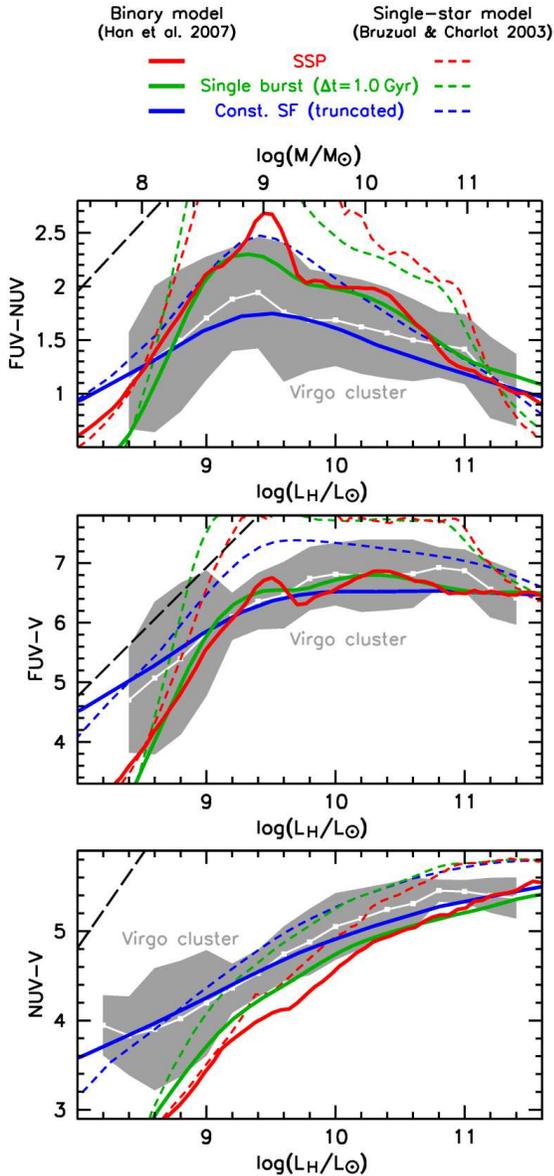}
  \caption{{\bf Models vs.\ observations.} 
Each panel shows the colors and near-infrared $H$ luminosities of Virgo
cluster early-type galaxies. Average color values (white dots) are
calculated within a luminosity interval of $\pm 0.3$ dex, and are given
every 0.2 dex, connected by the white line.
The grey shaded area represents
the RMS scatter. 
The data were extracted from
\citet{bos05} using the {\it Dexter} tool \citep{dexter}, and comprise
76 galaxies in FUV$-$NUV (top), 70 in FUV$-V$ (middle), and 97 in NUV$-V$
(bottom). The average number of galaxies contained in each bin is 13 (top
and middle) and 16 (bottom); bins with less than 5 galaxies are excluded.
The black dashed lines in the upper left of each panel mark the
detection limit given by \citet{bos05}.
The upper axis of the top panel gives the stellar mass that has been
determined from the $H$ luminosity using the SSP mass-to-light ratio evolution
of HPL07.
The time evolution of synthetic stellar populations is shown with the same
line styles as in Fig.~\ref{fig:noobs}. For these, the age ranges used for
plotting were chosen to provide 
the best visual match to the observed FUV$-$NUV colors, allowing age
steps of 0.05 dex. The
resulting age ranges for the HPL07 binary model are
$8.6\le\log(t_{\rm SSP}/{\rm yr})\le9.75$,
$8.1\le\log(t_{\rm Single}/{\rm yr})\le9.6$,
and $8.4\le\log(t_{\rm Const}/{\rm yr})\le9.4$, and for the single-star model
of \citet{bc03}, 
$8.6\le\log(t_{\rm SSP}/{\rm yr})\le10.0$,
$8.1\le\log(t_{\rm Single}/{\rm yr})\le9.95$,
and $8.35\le\log(t_{\rm Const}/{\rm yr})\le9.8$.
Note that the ``age'' $t$ denotes, for all model populations, the time
since the \emph{end} of star formation.
The given interval boundaries correspond to the low and high luminosity
end of the grey-shaded area in the upper panel, not to the boundaries of the
figure.
The same age ranges were used for the middle and bottom panel.
  }
  \label{fig:withobs}
\end{figure}

The {\it GALEX} UV colors of Virgo early-type galaxies were presented by
\citet{bos05} in relation to their near-infrared $H$ luminosity. A dividing
luminosity of dwarfs and giants \citep[as classified by][]{vcc} can be
roughly defined at $\log(L_H/L_\odot) \approx 9.6$ \citep[their Fig.~1]{bos05}. 
Here, we compare these data to synthetic colors from the HPL07 binary model,
as well as the BC03 single-star model. Our intention is to investigate whether
the models are able to simultaneously match the observed range in FUV$-$NUV,
FUV$-V$, and NUV$-V$, and whether the observed relation of color and
luminosity could be explained with an age variation only.

Fig.~\ref{fig:withobs} shows the observed colors as binned averages (white
line and dots), along with the respective RMS scatter (grey shaded
area). Colors were corrected for Galactic extinction by assuming an
average $E(B-V)=0.03$ \citep{gil07}, leading to $A_{\rm FUV}=A_{\rm
  NUV}=0.24, A_V=0.10$ \citep{sch98}. Color values are given 
in the AB magnitude system \citep{ABsystem}. No internal extinction correction
was applied: for the minor population of Virgo cluster early type
galaxies with ongoing central star formation, a typical value for $E(B-V)$
of 0.1 was found for their centers \citep{p2}, thus being much
smaller for the galaxies as a whole, and certainly negligible for the vast
majority of the sample.

For each synthetic stellar population shown in Fig.~\ref{fig:noobs}, we
selected an age range such that the time evolution of the model FUV$-$NUV color
traces best the relation with luminosity (top panel of
Fig.~\ref{fig:withobs}). The matching was performed visually by allowing age
steps of 0.05 dex. The age range was then kept fixed for the
other colors (middle and bottom panel).

We see from the top panel of Fig.~\ref{fig:withobs} that the SSP of HPL07
exceeds the observed color range at intermediate ages. This is no
surprise, though: since the reddest color of the SSP is reached during a
very brief period, any realistic
population -- with a finite duration of star formation rather than a
delta-peak -- can never become so red, since not all of its stars
reach this period at the same time. Indeed, the model population that formed
from a single burst of finite length lies within the observed color range,
although still 
somewhat offset from the mean values. The best match is provided by the model
population that formed through a constant star formation rate with
a varying truncation time, which lies relatively close to the observed
mean values.

In reality, we might expect that the most massive galaxies
did not only form earlier, but also needed less time to make their
stars. Therefore, one might prefer the SSP or single burst model at high
luminosities,
whereas in the dwarf regime, continuous and rather recently
truncated star formation might appear more plausible. In any case, the
binary model is able to explain the observed turnaround in FUV$-$NUV,
as well as the apparently different slopes of dwarfs and giants, in a
simple and straightforward way, by only varying the average stellar
population age. Moreover, the model with constant star formation is able to
simultaneously match the observed range of colors in FUV$-V$ and NUV$-V$. For
the SSP and the single burst population, no match is achieved in NUV$-V$
except for the high-mass end. Note that we are regarding the $H$ luminosity as
a proxy for the stellar mass of a galaxy. A transformation into mass is shown
as the uppermost abscissa in Fig.~\ref{fig:withobs}, using the mass-to-light
ratio in $H$ as given by the HPL07 SSP.

As for the BC03 models, both the SSP and the single-burst population
exceed the observed FUV$-$NUV color range by far. This is not only the case
for the comparison with
Virgo cluster galaxies: in fact, only three out of 874 galaxies of the
{\it GALEX} UV atlas of nearby galaxies \citep{gil07} are redder than a
FUV$-$NUV value of 2.6, whereas the BC03 models reach much redder values.
Nevertheless, the BC03 model colors for continuous star formation
just fall within the observed RMS scatter, and could thus provide a
similar interpretation as the HPL07 model. In contrast
to the binary model, though, none of the BC03 populations is able to provide a 
simultaneous match to all three colors. 

\subsection{Comparison to other models}
\label{sec:sub_othermodels}

\begin{figure}
  \epsscale{1.}
  \plotone{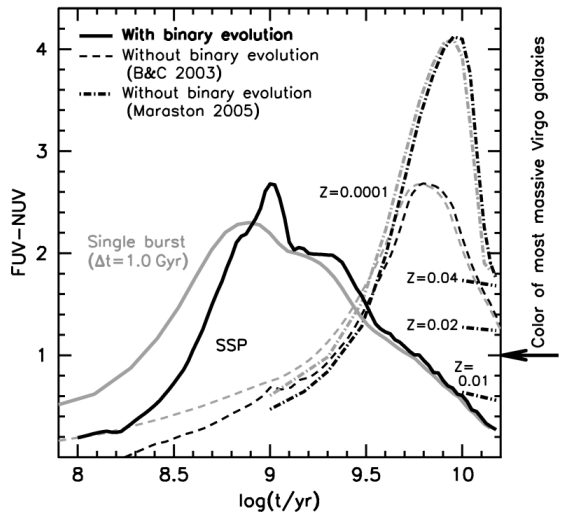}
  \caption{{\bf Alternative models.}
    Time evolution of FUV$-$NUV color of various synthetic stellar populations,
    computed with the binary model of \citet[solid lines]{han07} for solar
    metallicity ($Z=0.02$)
    and with the single-star model of \citet[dash-dotted lines]{mar05} with
    blue 
    horizontal branch morphology for metallicities $Z=0.0001$ (1/200th
    solar), $Z=0.01$ (1/2 solar), $Z=0.02$ (solar), and $Z=0.04$ (2 times
    solar). Note that for the models
    with $Z\geq 0.01$, \citet{mar05} only computed SSPs for the age of 10
    and 15\,Gyr.
    For comparison with the low-metallicity model of \citeauthor{mar05},
    we also show the \citet{bc03} model for $Z=0.0001$ (dashed lines).
    The FUV$-$NUV color reached by the brightest Virgo early-type galaxies
    is indicated by the arrow.
  }
  \label{fig:othermodels}
\end{figure}

The observed mass-to-light ratio evolution of early-type galaxies with
redshift has been found to be reproducible with the population synthesis
models of \citet{mar05}, but not with those of \citet{bc03}, possibly due
to the different AGB treatment \citep{vdW06}.
Moreover,
the \citeauthor{mar05} models have been shown by \citet{mar00} to match
the UV and optical spectra of a sample of early-type galaxies when a
metal-rich stellar population with red horizontal branch (HB)
morphology is combined with a small contribution of an old
metal-poor population with blue HB morphology. It therefore
appears worthwhile to compare these single-star models to the HPL07 binary
model. 

\citet{mar00} point out that, while the minor population of old metal-poor
stars is able to account for observed large values of $H_\beta$, it is the
dominant metal-rich population that reproduces the observed 1500$-V$
color. However, as can be seen from their Fig.~5, this metal-rich population
does not show an uprise of the spectral slope in the FUV, and does thus not
exhibit blue FUV$-$NUV colors, which are at the focus of our present
study. Only the metal-poor population exhibits such an uprise. We therefore
show in Fig.~\ref{fig:othermodels} the time evolution of the model of
\citet{mar05} with $Z=0.0001$ and blue HB morphology (using a
\citet{salpeter}
IMF), without
combining it with a metal-rich population of red HB morphology. It can be seen
from the figure that this model reaches FUV$-$NUV colors that are even redder
than the solar metallicity model of BC03 (Fig.~\ref{fig:noobs}), and also
redder than the BC03 model with 
$Z=0.0001$. Moreover, at large ages, the model is not able to account for the
blue FUV$-$NUV values that are observed for the Virgo cluster galaxies at the
high-mass end.

Since \citet{mar00} emphasized that \emph{metal-rich} populations with blue HB
morphologies could provide another alternative to explaining large  $H_\beta$
values without invoking young ages, \citet{mar05} computed models with blue HB
morphology also for high metallicities. These models, which were only computed
for ages of 10 and 15\,Gyr, are
shown in the figure as the short lines on the right-hand side. Only the
population with metallicity $Z=0.01$ reaches colors that are blue enough to
account for the observed ones. However, if it were to fully explain the
FUV$-$NUV colors of the most massive early-type galaxies, this population
would have to be the predominant one in any composite population, which
appears unlikely given its subsolar metallicity.

Another model that was shown to match the UV properties of early-type galaxies
is the single-star model of \citet{yi98}. While at old ages, its
FUV$-$NUV colors are blue enough to match the values of the Virgo
galaxies (as are the BC03 colors), at intermediate ages it
reaches comparably red colors in FUV$-$NUV and FUV$-$V as the BC03 model. It 
is thus also not able to match the observations with a single sequence like
the HPL07 model does.  See \citet{han07} for a 
discussion of this and other models that were proposed to explain the
UV-upturn.

\subsection{A possible color selection effect?}
\label{sec:sub_select}

\begin{figure}
  \epsscale{1.}
  \plotone{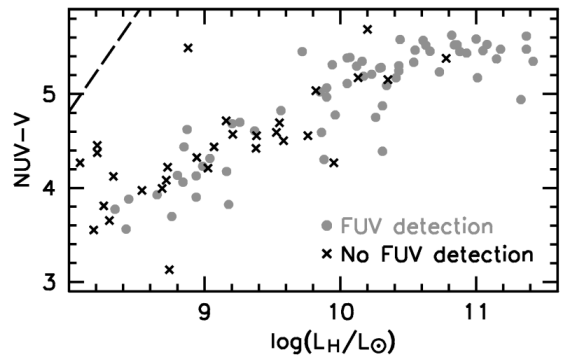}
  \caption{{\bf Assessing a possible color selection effect.}
    NUV$-V$ color versus near-infrared $H$ luminosity of Virgo
    cluster early-type galaxies, as given in \citet{bos05}. Grey circles
    denote objects that have a FUV detection, whereas black crosses are
    objects with no FUV detection.
    The black dashed line marks the
     detection limit given by \citet{bos05}.
  }
  \label{fig:select}
\end{figure}

The NUV and FUV photometry presented by \citet{bos05}, which we use
here, is only available for galaxies that were detected by {\it GALEX} in
NUV and FUV, respectively. This could cause a color selection bias, in the
sense that galaxies with redder FUV$-$NUV would be too faint in the FUV to be
detected. Since the sample of \citeauthor{bos05} becomes bluer in
FUV$-$NUV towards the faint end (Fig.~\ref{fig:withobs}), it is important
to assess whether this could be purely due to a selection effect.

In Fig.~\ref{fig:select} we show the NUV$-V$ color of all galaxies, as
given in \citet{bos05}. Those that are detected in the FUV, and therefore
enter the FUV$-$NUV and FUV$-V$ diagrams, are shown as grey circles; all
others are denoted by black crosses. The dwarf galaxy regime starts at
about $\log(L_H/L_\odot) \approx 9.6$, thus spanning 4 magnitudes
($\Delta\,\log(L_H/L_\odot) \approx 1.6$) in this sample. At the faintest
luminosities, a handful of galaxies are redder than the apparent mean
relation, and these are indeed not detected in the FUV. However, down to
$\log(L_H/L_\odot) \approx 8.4$, only a mild offset -- if at all
significant -- is seen between FUV-detections and non-detections, of the
order of $\sim 0.1$ mag.

If the UV emission is caused by a young stellar
population, then an offset of $\sim 0.1$ mag in
NUV$-V$ would correspond to an offset of not more than $\sim 0.15$ mag in
FUV$-$NUV. This can be seen from Fig.~\ref{fig:withobs} when considering
by how much the model colors change per luminosity interval.
This translation from one color
into the other is possible because a young stellar population
exhibits blue colors in both NUV$-V$ and FUV$-$NUV. For an old stellar
population showing a UV-upturn, this would of course be different --- but
the presence of a strong UV-upturn at these faint luminosities is neither
deduced in previous studies nor in this work. 

We summarize that the above investigation points towards a mild selection
effect only, which does not support the interpretation of the bluening of
FUV$-$NUV in the dwarf regime as an artificial feature. We also point out that only
a relatively small fraction of Virgo cluster early-type dwarfs was found
to exhibit ongoing star formation \citep{p2}; see
Sect.~\ref{sec:sub_dwarfgiant} for a discussion of this point.
Nevertheless, from Fig.~\ref{fig:withobs} it can be seen that the
detection limits are closest to the bulk of the data in FUV$-V$, where faint
red galaxies might be missed. This could indirectly lead to a selection
effect in FUV$-$NUV to some extent, even if the detection limits given for
FUV$-$NUV itself appear to be less disturbing.
Certainly a more comprehensive assessment of the limitations
of GALEX photometry for different stellar populations, which is
beyond the scope of the present study, would be necessary to obtain
precise estimates of the number and colors of potentially undetected galaxies.


\section{Discussion and summary}
\label{sec:discuss}

\subsection{The UV colors of early-type galaxies}

From an analysis of GALEX data for over 6000 galaxies with the (single-star)
models of \citet{bc03},
\citet{sal05} concluded that ``the star formation history of a galaxy can
be constrained on the basis of the NUV$-r$ color alone''. However, our
Fig.~\ref{fig:withobs} shows that FUV$-$NUV contains important information
on the stellar population of an early-type galaxy: by taking into account
the formation of hot subdwarf stars that contribute significantly to the
FUV flux, the binary model of \citet{han07} can easily explain
the observed FUV$-$NUV colors of Virgo cluster early-type galaxies with a
younger stellar population age at lower luminosities.
This continuous sequence in luminosity -- and hence, galaxy mass -- 
also provides a natural interpretation of the observed anticorrelation between
FUV$-$NUV and $B-V$ for nearby early-type galaxies \citep{don07}.
Similarly, the findings of \citet{ric05}, that ``quiescent''
early-type galaxies are bluer in FUV$-$NUV than ``star-forming''
early types, can be straightforwardly explained with the latter having
formed the majority of their stars later than the former.

It needs to be stressed that the dependence of FUV$-$NUV on the
stellar population 
age differs from the behavior of the actual UV-upturn, depending on the
definition of the latter.
If the UV-upturn is defined as the 
slope $\beta$ of the FUV spectrum, it remains at an almost constant
value soon after the onset of the formation of hot subdwarfs in a galaxy,
and does not change with age \citep{han07}. Consequently, different
definitions of the UV-upturn in the literature need to be carefully
distinguished. For example, \citet{ric05} adopt FUV$-r$ color as the
UV-upturn -- similar to the 1550$-V$ color analyzed by \citet{bur88} --
for which the corresponding age evolution of the binary 
model is very different from that of both $\beta$ and FUV$-$NUV
(Fig.~\ref{fig:withobs}; \citealt{han07}).

Our findings do not necessarily contradict the claims of recent
star formation in a significant fraction of early types
\citep{kav07,sch07}: these and similar studies mainly used solely the NUV
flux as indicator for the presence of young stars \citep[also
  see][]{yi05}. It is  important to point out that these studies primarily
focus on whether a \emph{small fraction} of young stars is present,
whereas we have considered the average age of the \emph{whole} stellar
population of a galaxy. Therefore, not surprisingly, the statement of
\citet{kav07} that galaxies with NUV$-r<5.5$ (roughly NUV$-V<5.2$) are
very likely to have experienced \emph{some} recent star formation
translates into a younger \emph{overall} stellar age in our study (cf.\
Fig.~\ref{fig:withobs}, bottom).

\subsection{Residual or recent star formation in dwarfs?}
\label{sec:sub_residual}

While the degeneracy between a somewhat younger mean age 
or a small fraction of young stars on top of an old population is
unavoidable to some extent, we would like to emphasize the difference
between \emph{recent} and \emph{residual} star formation. In
Fig.~\ref{fig:withobs}, the time since the end of star formation for the
binary model with constant star formation rate ranges from
$\log(t/{\rm yr})=8.4$ at the low-mass end to $\log(t/{\rm yr})=9.4$ at
the high-mass end. Therefore, when defining 
``recent star formation'' as having some fraction of stars younger than
1\,Gyr \citep[cf.][]{kav07, sch07}, all galaxies with
$\log(L_H/L_\odot) \lesssim 10$ would count in this scenario as objects
that experienced recent star formation. In contrast, the term ``residual
star formation'' \citep[cf.][]{bos05,bos08} refers to a small amount of
\emph{ongoing} star formation activity, which should not be confused with
the former.

VCC~1499, a morphologically early-type-like object \citep{vcc,p2} but having
post-starburst characteristics \citep{gav01}, is 
significantly bluer in FUV$-$NUV than the average value at
its luminosity \citep{bos05}, with the offset being about 2 times the RMS color
scatter. \citet{p2}, who identified Virgo early-type dwarfs 
with blue central $g-i$ colors, showed that the inner colors
of VCC~1499 are bluer than those of all early-type dwarfs having a blue
center. In that study, 
the weakest blue centers that could still be
identified as such
 exhibit a $g-i$ color difference of 0.1 mag between inner
and outer galaxy regions, whereas the value for VCC~1499 is 0.5 mag. VCC~1499
thus stands out at least as clear in the optical Sloan Digital
Sky Survey data \citep{sdssdr5} as it does in the {\it GALEX} data, partly
because both the signal-to-noise ratio and the resolution are much better
in the former. It is therefore not necessarily the case that residual or
very recent star formation activity can be identified down to lower levels
with {\it GALEX} than with optical data.

In the sample of \citet{p2}, VCC~1499 is the only
galaxy for which the population synthesis
model fits to the spectrum of the very central region yielded a mass
fraction of young stars ($<500$\,Myr) of more than 10\%. The other
early-type dwarfs with a blue central region -- of which several show
$H_\alpha$ emission from ongoing star formation \citep{p2,bos08} --
were deduced to have several thousandths of their mass in very young stars
($<10$\,Myr), and several percent of their mass in stars younger than
500\,Myr. These objects make up $\sim 15\%$ of early-type dwarfs among
the brightest two magnitudes. We would thus expect that the majority of
Virgo early-type dwarfs has a
lower fraction of young stars that could not be identified by
\citet{p2}. Let us therefore consider the age ranges that were adopted in
Fig.~\ref{fig:withobs} (specified in the caption).
For the SSP and the single burst model, more than one
third of the stars were formed less than 500 Myr ago at low galaxy
masses. If this was the real situation, \citet{p2} should have detected
recent star formation like in VCC~1499 in the vast majority of early-type
dwarfs. However, both the SSP and the single burst model do \emph{not}
match the NUV$-V$ color (Fig.~\ref{fig:withobs}, bottom), for the very
same reason: in case of significant recent star formation, NUV$-V$ would
be much bluer than what is actually observed. (VCC~1499 does have such a very
blue NUV$-V$ color). Only the model with constant star formation
is able to match the observed colors at the faint end. For this model,
only 2\% of all stars formed less than 500 Myr ago, which, given the above
considerations, would indeed not have been detected by \citet{p2}, leading
to a consistent picture.

This scenario is also in accordance with the indications for just a small
color selection effect in FUV$-$NUV (see Sect.~\ref{sec:sub_select}), which
should be much larger if most of the FUV-detected galaxies contained a
significant amount of young stars.
Furthermore,
we would like to remark that
only a very small fraction of early-type galaxies in the Virgo cluster contain a
significant amount of neutral hydrogen \citep{diSer07,gav08}, again in agreement
with the above picture.

\subsection{From dwarfs to giants}
\label{sec:sub_dwarfgiant}

Do dwarf and giant early-type galaxies form one continuous parameter
sequence, despite the rather different formation scenarios suggested for
them \citep[e.g.][]{bell06,bos08}? Their overall shapes follow a structural
continuum with luminosity \citep{jer97,gra03,gav05a}. In nearby galaxy
clusters, early-type dwarfs were reported to follow the
color-magnitude relation (CMR) of giant early types, at least on average
(\citealt{sec97, conIII}; also see \citealt{and06}).
\citet{smi08} recently presented a
well-defined optical CMR of early-type dwarf and giant galaxies in the
Antlia cluster, which shows no change in slope within a range of 9
magnitudes. 
However, from {\it GALEX} UV colors, a pronounced dichotomy seemed
to be present between Virgo cluster dwarf and giant early types
\citep{bos05}, hinting at systematic differences in their star formation
histories. With our above analysis, which takes into account the FUV flux
of hot subdwarf stars, we have shown that this color
behavior can be naturally explained with a continuous sequence
 of longer duration and later truncation of star formation at lower galaxy
masses, all the way from giants to dwarfs. These results do point towards
a fraction of relatively young stars in early-type dwarfs (see
Sect.~\ref{sec:sub_residual}); however, residual star formation activity,
as suggested by \citet{bos05,bos08}, is not necessary to account for the
observed colors.
While a number of Virgo 
cluster early-type dwarfs with central star formation activity are known
\citep{p2}, these only constitute a minor subpopulation \citep{p3}.

A continuous sequence of longer star formation duration at lower
galaxy masses has been found by \citet{tho05} for giant early-type
galaxies, in agreement with the larger mass fraction of young stars at
lower galaxy masses reported by \citet{fersilk00}. Given the
apparent continuity of the optical CMR down to early-type dwarfs, and the
reproduction of the turnaround in FUV$-$NUV by the HPL07 model,
it appears reasonable that this
trend in the star formation history might simply be continued to the
dwarfs. Along these lines, a clear relation of longer duration of star
formation at lower $H$ luminosity was found by \citet[their
  Fig.~11]{gav02} in their 
spectrophotometric study of Virgo cluster early-type dwarfs and giants.
A continuous star formation with rather late truncation for dwarfs would
also be in agreement
with the study of \citet{bos08}, who compared multiwavelength observations
of Virgo cluster dwarf galaxies to chemo-spectrophotometric models and
concluded that the majority of early-type dwarfs could have been formed
recently through ram-pressure stripping of late-type galaxies.

How plausible would a continuity between the star formation histories of
giant and dwarf early-type galaxies appear in the light of the recent
infall of the dwarf progenitors -- following the scenario proposed by
\citet{bos08} -- as opposed to the dynamically relaxed and centrally
concentrated population of giant early types \citep[cf.][]{conI}?
The fact that for giant early-type galaxies, significant stellar
population differences are found between clusters and the field
\citep[e.g.][]{tho05}, but not between different clusters
\citep[e.g.][]{bow92,and03}, indicates that the environmental density played an
important role already at early epochs. Would we thus expect a continuity
of cluster giants and dwarfs if the progenitors of the latter just recently
arrived from a low-density environment? One key to this question might be
in the recent identification of several early-type dwarf subclasses that
have significantly different shapes, colors, and/or spatial distributions
within the Virgo cluster \citep{p3}. Early-type dwarfs with weak disk
features \citep{p1}, with blue central regions \citep{p2}, as well as
those without a bright stellar nucleus all have a rather flat shape and
populate a similar density regime as the late-type cluster galaxies
do.
Only the ordinary nucleated early-type dwarfs fit into the classical image
of dwarf elliptical galaxies, in that they have a spheroidal shape and are
concentrated towards the cluster center \citep{p3}. Moreover, they exhibit
somewhat older and/or more metal-rich stellar populations than the other
subclasses.
 This diversity renders a common formation mechanism
for all early-type dwarfs highly unlikely \citep{p4}.
One can thus speculate whether those nucleated early-type dwarfs have
resided in the cluster since a long time already \citep{oh00}, along with
the giant early types, 
and might therefore be responsible for the above picture of a
continuum from giants to dwarfs --- after all, they constitute the
majority of the early-type dwarf population. The other dwarf subclasses
might be those that formed through transformation of infalling galaxies,
as suggested by \citet{bos08}.
Future studies need to investigate in more detail the characteristics of
the different subclasses of early-type dwarfs, as well as their relation
to giant early types, in order to gain further insight into the physical
mechanisms responsible for their formation.

While still being relatively simple, the binary model of \citet{han07} is
able to account for the observed range of FUV$-$NUV color of Virgo cluster
early-type galaxies, and provides a natural explanation for its
behavior with luminosity. We thus believe that it will become an important
tool in future studies of the ultraviolet light from galaxies. Further
development of the model, such as an extension to various metallicities,
is in progress. Moreover, the diagnostic UV-optical color-color diagrams
presented by \citet{han07} enable us to study
the star formation histories of early-type galaxies 
in a more detailed way than presented here, which will be subject of a
forthcoming paper.


\acknowledgements
    We thank the referee for constructive suggestions that helped
    improving the manuscript, and Philipp
    Podsiadlowski for stimulating discussions.
    T.L.\ is supported within the framework of the Excellence Initiative
    by the German Research Foundation (DFG) through the Heidelberg
    Graduate School of Fundamental Physics (grant number GSC 129/1).
    Z.H.\ acknowledges the support from the Natural Science Foundation of China
   (Grant Nos 10433030, 10521001 and 2007CB815406).
    This research has made use of NASA's Astrophysics Data
    System Bibliographic Services.

 



\end{document}